\documentclass{aa}

\usepackage{graphicx}
\usepackage{enumerate}
\usepackage{txfonts}
\usepackage{url}
\usepackage{wasysym}
\usepackage{longtable}
\usepackage{array}
\usepackage{tikz}
\usepackage[pdftitle={Signal preservation of exomoon transits during light curve folding}, colorlinks = true, breaklinks = true, citecolor = blue, linkcolor = blue, urlcolor = blue, pdfauthor = {Ren\'{e} Heller}]{hyperref}
\usepackage{esvect}  

\begin{document} 



\title{Signal preservation of exomoon transits during light curve folding}
\titlerunning{Signal preservation of exomoon transits during light curve folding}

\author{
  Ren{\'e} Heller\inst{1,2}
  \and
  Michael Hippke\inst{3,4}
}

\institute{
   Max-Planck-Institut f\"ur Sonnensystemforschung, Justus-von-Liebig-Weg 3, 37077 G\"ottingen, Germany\\ \email{heller@mps.mpg.de}
   \and
   Institut f\"ur Astrophysik, Georg-August-Universit\"at G\"ottingen, Friedrich-Hund-Platz 1, 37077 G\"ottingen, Germany
   \and
   Sonneberg Observatory, Sternwartestra{\ss}e 32, 96515 Sonneberg, Germany\\
   \email{michael@hippke.org}
   \and
   Visiting Scholar, Breakthrough Listen Group, Berkeley SETI Research Center, Astronomy Department, UC Berkeley
}

   \date{Received 9 October 2021; Accepted 4 November 2021}

 
\abstract{
In the search for moons around extrasolar planets (exomoons), astronomers are confronted with a stunning observation. Although 3400 of the 4500 exoplanets were discovered with the transit method and although there are well over 25 times as many moons than planets known in the Solar System (two of which are larger than Mercury), no exomoon has been discovered to date. In the search for exoplanet transits, stellar light curves are usually phase-folded over a range of trial epochs and periods. This approach, however, is not applicable in a straightforward manner to exomoons. Planet-moon transits either have to be modeled in great detail (including their orbital dynamics, mutual eclipses, etc.), which is computationally expensive, or key simplifications have to be assumed in the modeling. One such simplification is to search for moon transits outside of the planetary transits. The question we address in this report is how much in-transit data of an exomoon remains uncontaminated by the near-simultaneous transits of its host planet. We develop an analytical framework based on the probability density of the sky-projected apparent position of an exomoon relative to its planet and test our results with a numerical planet-moon transit simulator. For exomoons with planet-moon orbital separations similar to the Galilean moons, we find that only a small fraction of their in-transit data is uncontaminated by planetary transits: 14\,\% for Io, 20\,\% for Europa, 42\,\% for Ganymede, and 73\,\% for Callisto. The signal-to-noise ratio (S/N) of an out-of-planetary-transit folding technique is reduced compared to a full photodynamical model to about 38\,\% (Io), 45\,\% (Europa), 65\,\% (Ganymede), and 85\,\% (Callisto), respectively. For the Earth's Moon, we find an uncontaminated data fraction of typically just 18\,\% and a resulting S/N reduction to 42\,\%. These values are astonishingly small and suggest that the gain in speed for any exomoon transit search algorithm that ignores the planetary in-transit data comes at the heavy price of losing a substantial fraction of what is supposedly a tiny signal in the first place. We conclude that photodynamical modeling of the entire light curve has substantial, and possibly essential, advantages over folding techniques of exomoon transits outside the planetary transits, in particular for small exomoons comparable to those of the Solar System.}


   \maketitle
%

\section{Introduction}

Folding 
stellar light curves has become a standard when searching for exoplanetary transits \citep{2002A&A...391..369K,2019A&A...623A..39H}. A similar technique has recently been proposed for exomoons \citep[][K21]{2021MNRAS.507.4120K} in cases where the host planet shows transit timing variations due to the gravitational interaction with its satellite \citep{1999A&AS..134..553S}. A key concern for such an exomoon folding algorithm is that it has to work in the presence of the planetary transit, which is much deeper than the moon transit. One way around this is to search for photometric signatures of a transiting exomoon in the phase-folded light curve of the planet \citep{2012MNRAS.419..164S}. The resulting transit feature of the moon is a smeared version of the moon's individual transits because phase-folding with respect to the planetary transits adds moon transits from a range of possible planet-moon orbital configurations \citep{2014ApJ...787...14H}.

If, however, the aim is to separate the moon's transit signature from the planetary transits, ``a clever dynamical phase-folding would be necessary to recover a signal'', as splendidly put by \citet{2017MNRAS.467.1694M}. K21 suggests masking out the planetary transits and searching for any residual in-transit data of the moon before and after the planetary transits, all of which might still yield a detectable moon transit when phase-folded on the correct orbital ephemerides of the moon. One example system of a Sun-like star, a Jupiter-sized exoplanet with an orbital period of 60\,d (orbital semi-major axis of 0.3\,AU), and an exomoon in an extremely wide orbit (at the Hill radius) around this planet was simulated to demonstrate the capabilities of a transit folding algorithm for exomoons. This system was tailored to allow for a sufficiently large number of transits within four years of simulated data, similar to the baseline of the Kepler primary mission. The folding algorithm was found to retrieve the injected signal with a signal-to-noise ratio (S/N) of 82\,\% of the S/N obtained with a full photodynamical planet-moon transit model.

Here we are interested in a general solution to the problem of an exomoon's in-transit data in the light curve that is uncontaminated by the planetary transits. Our aim is to answer the question if a folding technique for light curves could also work for exomoons in orbits similar to those of the Solar System moons.

\begin{figure}
    \centering
    \includegraphics[width=0.98\linewidth]{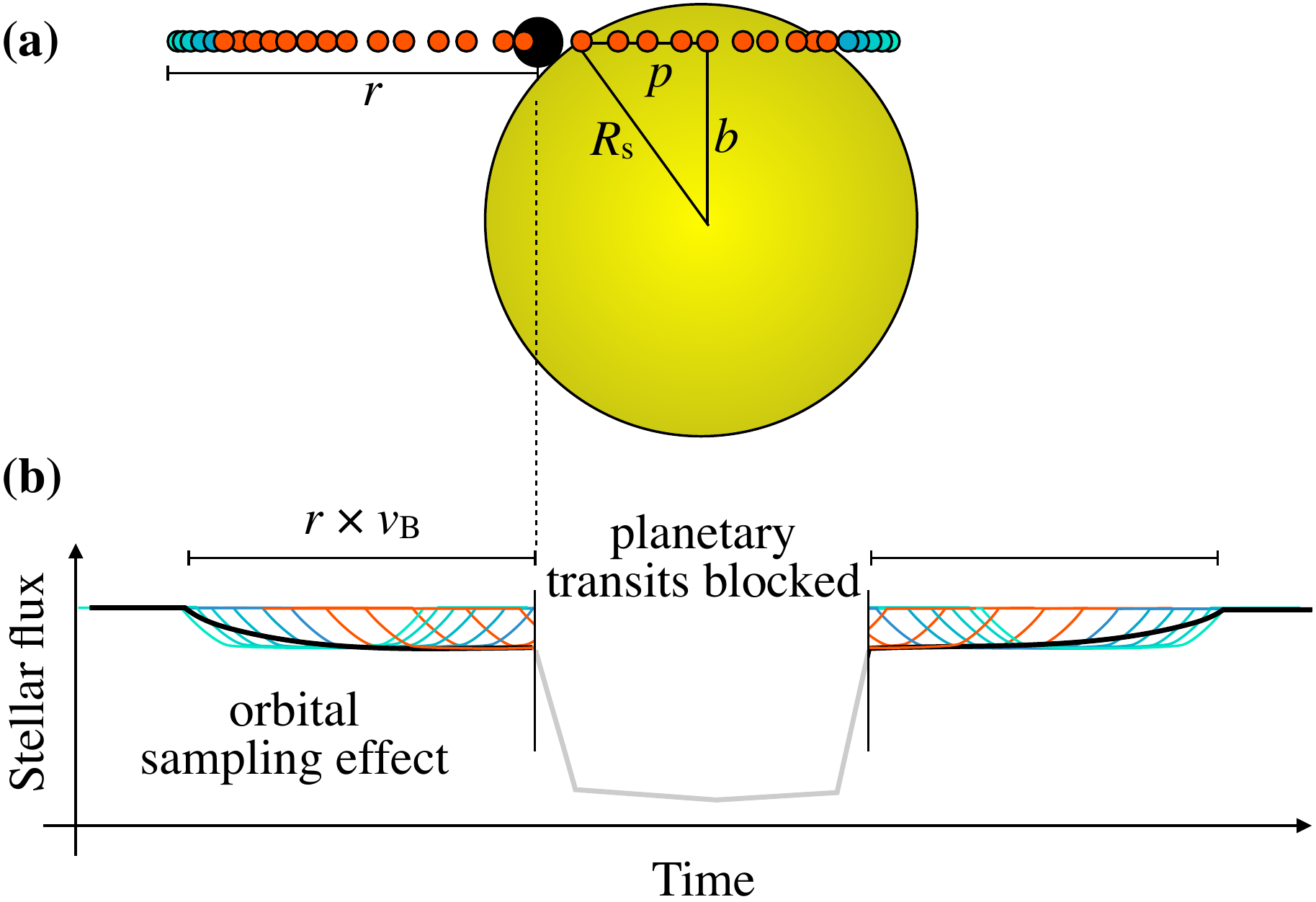}
    \caption{{\bf (a)} Transit geometry of the three-body system consisting of a star (large yellow circle), a planet (black circle), and a moon (orange and blue circles). The transit path across the star equals $2p = 2R_{\rm s}^2\sqrt{1-b^2}$. {\bf (b)} Illustration of an exomoon's orbital sampling effect (OSE) \citep{2014ApJ...787...14H,2016ApJ...820...88H} in the light curve. The circumstellar orbital velocity of the barycenter ($v_{\rm B}$) multiplied by the planet-moon orbital separation ($r$) defines the duration of the OSE feature before and after the planetary transit.}
    \label{fig:transits}
\end{figure}


\begin{figure}
    \centering
    \includegraphics[width=0.77\linewidth]{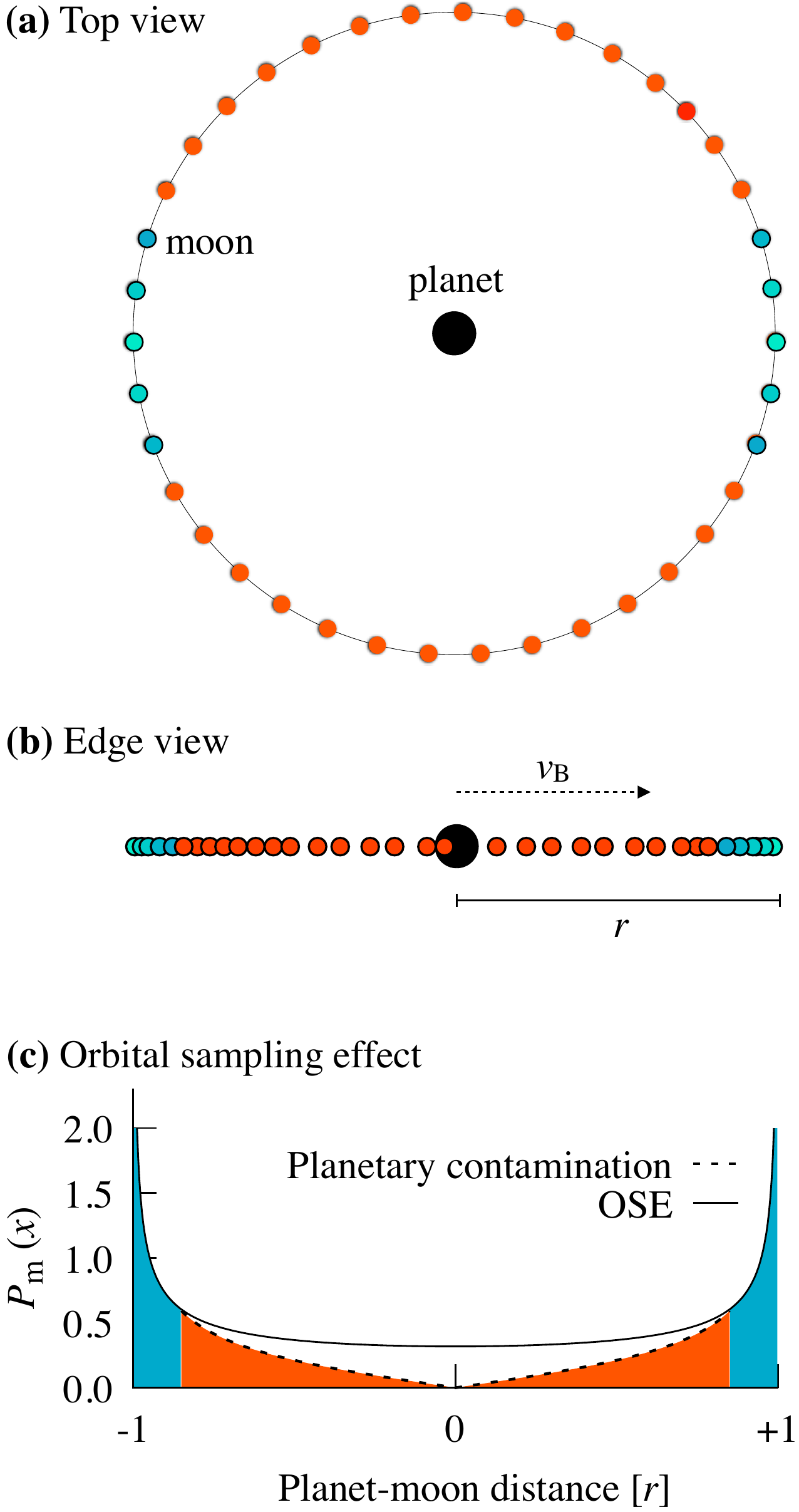}
    \caption{Orbital sampling effect of a moon around its host planet with the planet in the center of reference. {\bf (a)} In a circular orbit, the probability to find the moon at any given location along its orbital path is equal everywhere. If the moon's position were sampled many times, the samples would distribute homogeneously along its orbit. {\bf (b)} As a result of a projection effect, however, the samples would pile up toward the edges of the moon orbit if seen from the side. If the moon orbit is wider than the transit path across the star (blue moons), then the corresponding moon transits can be unaffected by planetary transits. {\bf (c)} If the moon exhibits sufficiently wide deflection from the planet, the corresponding in-transit data can fully contribute to the fraction of the moon's phase-folded in-transit data in the light curve (blue area). Moon transits closer to the planet are increasingly more contaminated the closer the moon is located to the planet. The curve assumes that the transit chord equals 85\,\% of the planet-moon separation ($2p/r=0.85$), a stellar radius of 50\,\% of the planet-moon separation, and an impact parameter $b=0.5$.}
    \label{fig:OSE}
\end{figure}

\begin{figure*}
    \centering
    \includegraphics[width=0.49\linewidth]{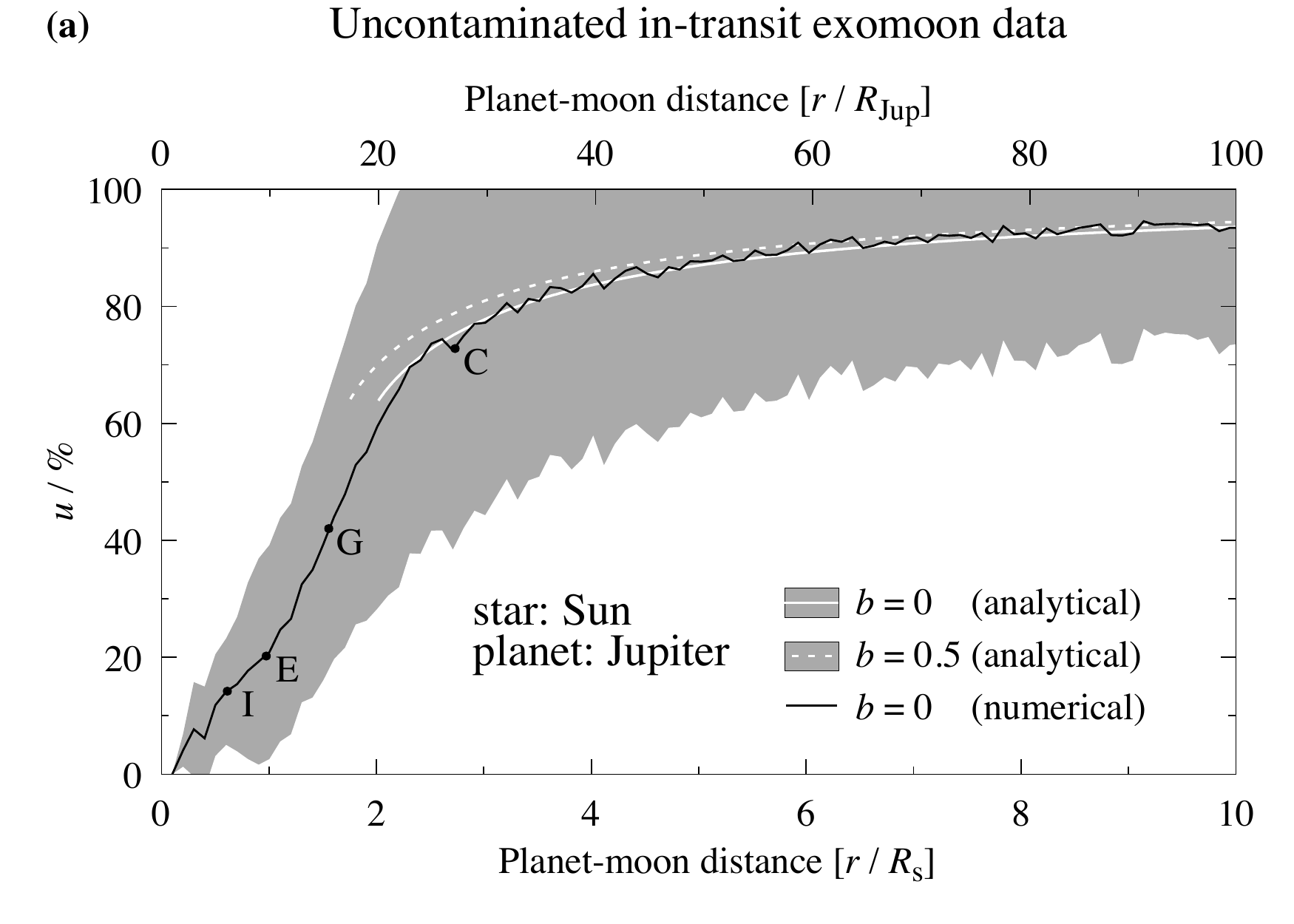}
    \hspace{0.15cm}
    \includegraphics[width=0.49\linewidth]{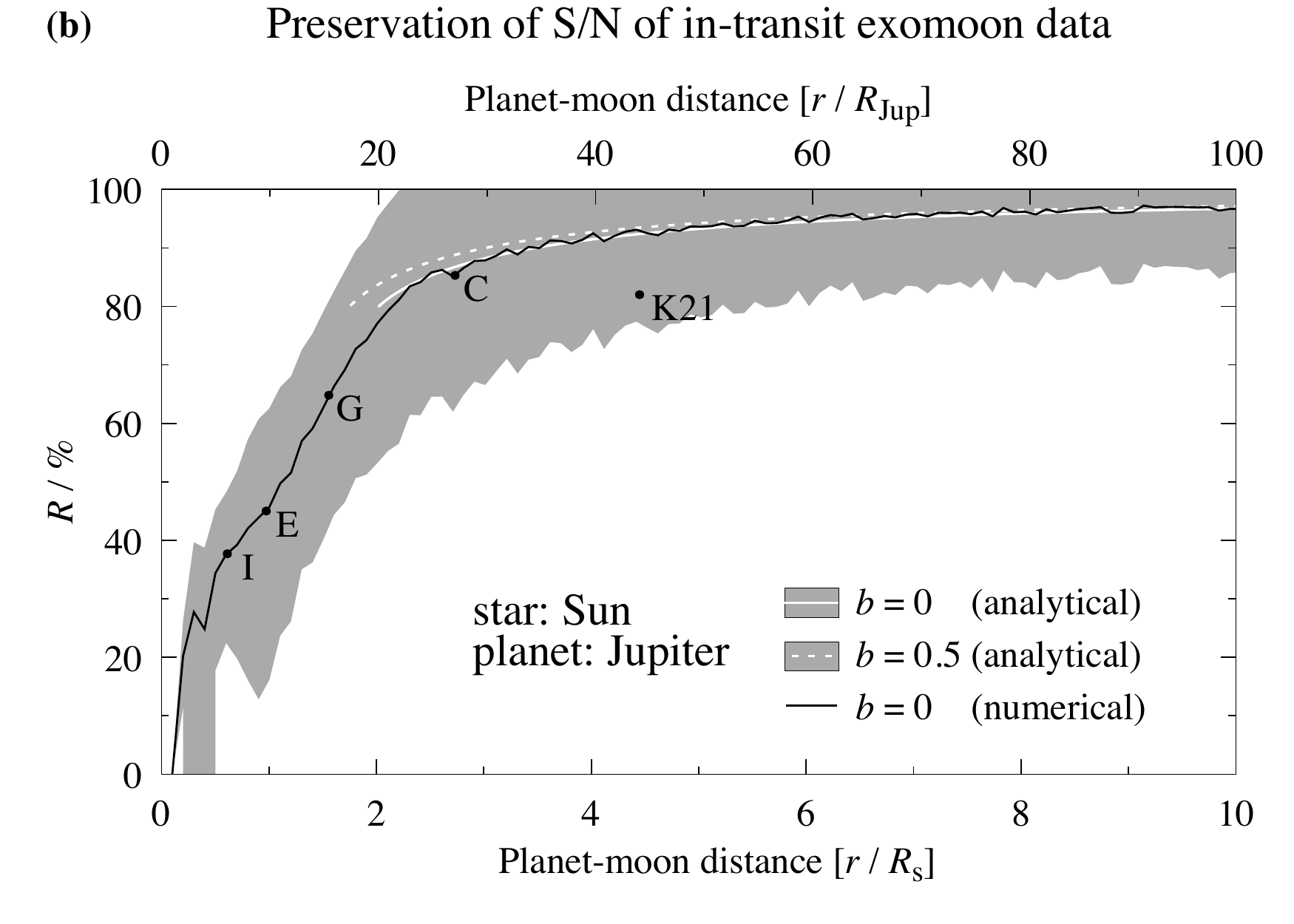}
    \caption{Contamination of exomoon transits by transits of the host planet. In the numerical simulations, a Jupiter-sized planet with a Ganymede-sized test moon was simulated to transit a Sun-like star with $b=0$. {\bf (a)} Percentage of an exomoon's in-transit data that is not contaminated by planetary transits. The black curve shows the mean value of $u$ after 10,000 numerical simulations for each planet-moon distance. The gray shading represents the standard deviation. The white solid and white dotted lines illustrate the analytical expressions as per Eq.~\eqref{eq:u_solved} for $b=0$ and $b=0.5$, respectively. {\bf (b)} The square root of the data shown in panel (a) estimates the S/N preservation of the exomoon transits upon rejecting any data during planetary transits. The measurement by K21 for a simulated system with 24 transits (using $b=0.5$) is labeled ``K21''.}
    \label{fig:u_SNR_Jup}
\end{figure*}

\section{Contamination of moon transits from the planet}
\label{app:contamination}

We consider a system of a star, a planet, and a moon. The orbits of the planet-moon barycenter around the star and of the planet and moon around their common barycenter are both circular and coplanar. Our aim is to estimate the fraction of the moon's in-transit data that is uncontaminated by planetary transits. We consider a system with an arbitrary number of transits. Of course, any observed system exhibits a limited number of transits in practice, and the fraction of the moon's uncontaminated in-transit data can, by chance, be very high (near 100\,\%) or very low (near 0\,\%), in particular if the number of transits is small. 

We devised two different approaches to investigate the contamination of exomoon transits by the planetary transits. One approach is a purely analytical framework and the other approach involves numerical validations of the predictions.

\subsection{Analytical framework}

Figure~\ref{fig:transits} shows the transit geometry of the planet and its moon in front of the star. The large number of moons in panel (a) is an illustration of the large number of transits that we consider in panel (b), where each moon corresponds to one transit of the moon (and its planet). All moon transits are centered around the mid-point of the planetary transit.


In this particular example, we chose the distance between the planet and its satellite ($r$) to be somewhat larger than the length of the transit chord, which is $2p = 2R_{\rm s} \sqrt{1-b^2}$ with $R_{\rm s}$ as the stellar radius and $b$ as the transit impact parameter. In this case, and if the motion of the moon and the planet are neglected for now, some of the sample moons (shown in blue) exhibit full transits before and after the planetary transit. Orange moons in Fig.~\ref{fig:transits}(a) and orange light curves in Fig.~\ref{fig:transits}(b) refer to any moon transit that is contaminated by the planetary transit. We are interested in the fraction of uncontaminated exomoon transits or uncontaminated in-transit data after many transits ($u$), that is, the fraction of blue moons in Fig.~\ref{fig:transits}(a) or blue transits in Fig.~\ref{fig:transits}(b).

We derived an analytical expression for $u$ for moon orbits that are wider than the transit chord. If $r > 2p$, the exomoon spends a certain fraction of some of its transits without contamination from the planetary transits. To compute this fraction of the exomoon's uncontaminated in-transit data, we considered the probability density of the moon along its projected orbit. Figure~\ref{fig:OSE} illustrates the interpretation of the probability density from a geometry perspective. Panel (a) shows a top-down view on the moon orbit around the planet, panel (b) shows an edge-on perspective as it can be expected during transit for coplanar orbits, and panel (c) shows the corresponding probability density \citep{2014ApJ...787...14H},

\begin{equation}\label{eq:ProbDens}
P_\mathrm{m}(x) = \frac{1}{{\pi}r\sqrt{\displaystyle 1 - \left(\frac{x}{r}\right)^2 } } \ \ .\end{equation}

\noindent
The integral of this function within the limits $-1 \leq r \leq +1$ equals 1 to ensure that the moon's probability to show up somewhere is normalized. From Fig.~\ref{fig:OSE}(c), we understand that any moon transits without planetary contamination (for $2p \leq |x| \leq r$) contribute according to the corresponding integral of $P_{\rm m}(x)$ to the fraction of the moon's uncontaminated in-transit data.

Whenever the moon cannot complete its transit before the planet enters the stellar disk (or begin its transit after the planet has left the stellar disk), a fraction of the moon transit is contaminated by the planetary transit. We find this fraction to be $x/(2p)$, which is equal to the deflection of the moon from the planet in transit as a fraction of the transit chord. As a result, these contaminated moon transits contribute an integral over $x/(2p) \times P_{\rm m}(x)$ of uncontaminated in-transit moon data.

\begin{figure*}
    \centering
    \includegraphics[width=0.49\linewidth]{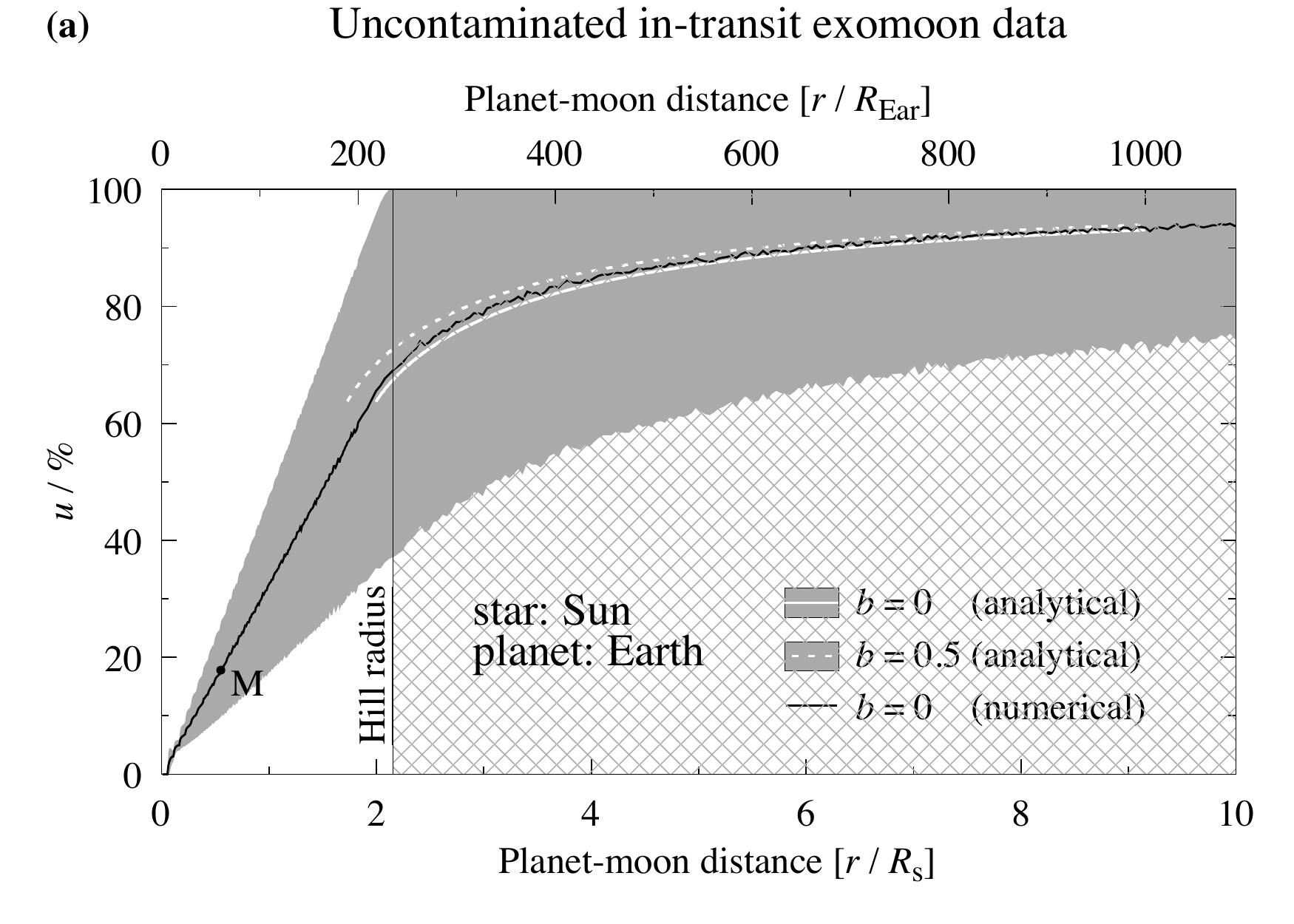}
    \hspace{0.15cm}
    \includegraphics[width=0.49\linewidth]{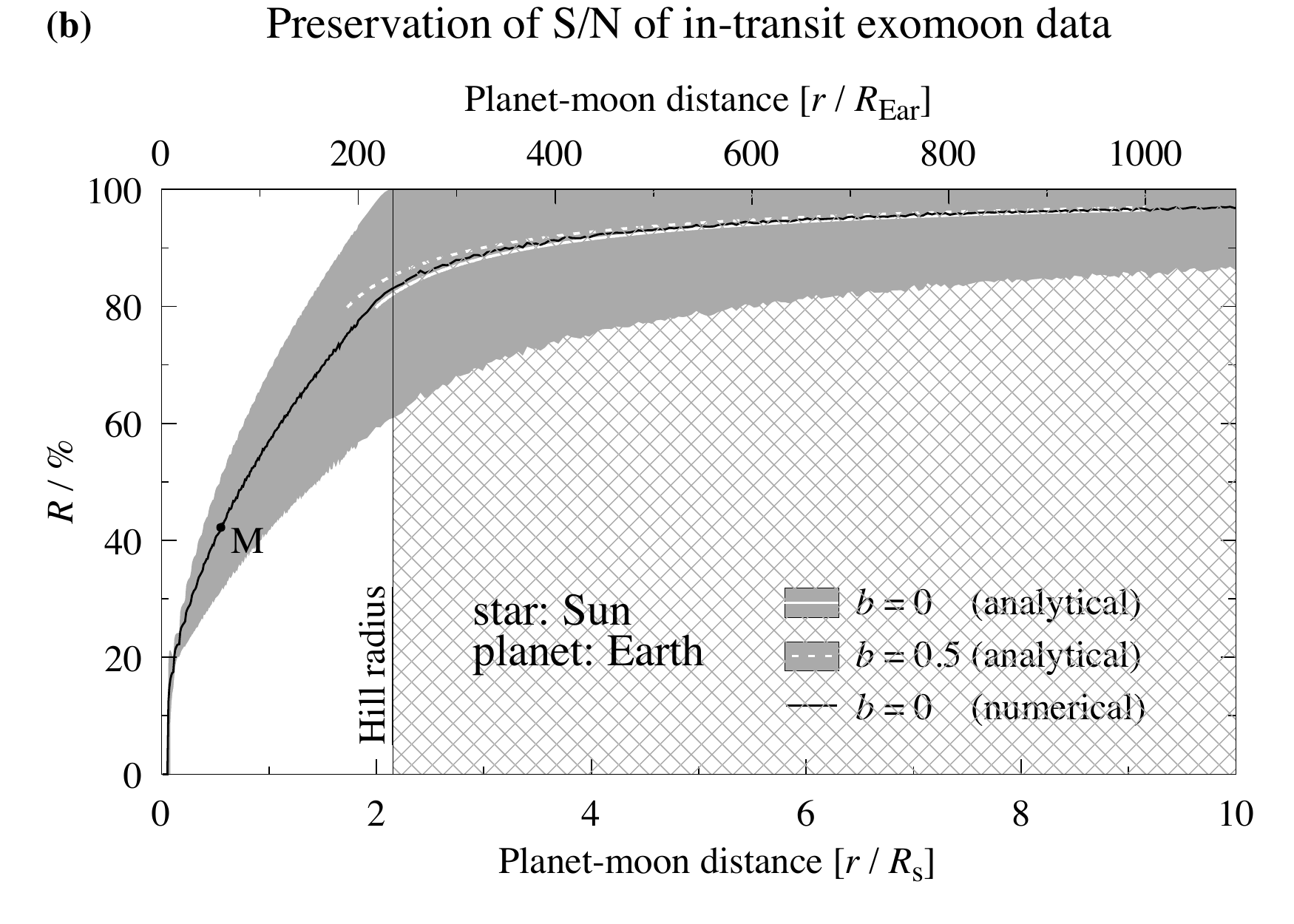}
    \caption{Same as Fig.~\ref{fig:u_SNR_Jup}, but now assuming an Earth-sized planet around a Sun-like star. For comparison, the orbital distance of the Moon around the Earth is indicated with an ``M''. The Earth's Hill radius is located at 235 Earth radii, or 2.15 solar radii, beyond which moons are dynamically unstable. Prograde moons are unstable beyond about half the Hill radius. The analytical solutions were truncated at 218 Earth radii for $b=0$ (solid white line) and at 189 Earth radii for $b=0.5$ (white dashed line), respectively.}
    \label{fig:u_SNR_Ear}
\end{figure*}

All things combined,

\begin{align} \label{eq:u_integral}
u =& \, 2 \int_{2p}^{r} \mathrm{d}x \ P_{\rm m}(x) + 2 \int_{R_{\rm p}}^{2p} \mathrm{d}x \ \frac{x}{2p} P_{\rm m}(x) \\ \nonumber
& \hspace{5.8cm} {\rm if} \ r > 2p \ ,
\end{align}

\noindent
where the factor of 2 accounts for the two symmetric wings of the probability density for positive and negative values of $x$. Also, the lower integration boundary in the second integral is set to the planetary radius ($R_{\rm p}$) to account for the loss of moon signal during planet-moon eclipses. For $x < R_{\rm p}$, exomoon transits cannot be folded, whereas a photodynamical transit model can be constructed to use planet-moon eclipses as genuine signal. The integrals in Eq.~\eqref{eq:u_integral} can be solved in closed form as
follows:
\begin{align} \label{eq:u_solved}
u =& \frac{2}{\pi} {\Bigg [} \frac{\pi}{2} - \arcsin{\Bigg (} \frac{2p}{r} {\Bigg )} + \frac{r}{2p} {\Bigg (} \sqrt{1 - \left(\frac{R_{\rm p}}{r} \right)^2} - \sqrt{1 - \left( \frac{2p}{r} \right)^2} {\Bigg )}  {\Bigg ]} \\ \nonumber
& \hspace{5.8cm} {\rm if} \ r > 2p \ .
\end{align}

\noindent
As an aside, closer examination of Eq.~\eqref{eq:u_solved} shows that $u$ increases when $b$ increases. That is to say, the closer the transit chord is to the stellar limb, the more of the moon's in-transit data can be preserved despite the rejection of any data that is contaminated by planetary transits. One must keep in mind though that the transit duration of both the planet and the moon decreases toward the limb proportionally to $(1-b^2)^{1/2}$ so that the moon signal effectively decreases as per $(1-b^2)^{1/4}$.

In different scenarios, where the moon orbit is smaller than the transit chord ($r < 2p$), some fraction of the moon's in-transit time is nevertheless uncontaminated. We estimate this fraction numerically in the following section. Future work could try to derive an analytical solution for this case by averaging over the moon's orbital velocity around the planet-moon barycenter.

To estimate the reduced S/N ($R$) of the moon's in-transit data after removal of any data that is contaminated by planetary transits, we note that ${\rm S/N} \propto 1/\sqrt{N}$, where $N$ is the number of data points. In the case of uniform time sampling, $N \propto u$ and thus $R \propto \sqrt{u}$.




\subsection{Numerical validation}

To test our analytical framework, we employed our publicly available\footnote{\href{https://github.com/hippke/PyOSE}{https://github.com/hippke/PyOSE}} planet-moon transit simulator {\tt PyOSE} \citep{2016ApJ...820...88H,2016A&A...591A..67H}. In brief, the moon's orbit around the planet is modeled with all six Keplerian elements, but neglecting the motion of the planet around the planet-moon barycenter. This latter effect is irrelevant for our purposes since we are interested in the moon's position with respect to the planet, not with respect to the planet-moon barycenter. Although not key to our subsequent analysis, mutual planet-moon eclipses are modeled in {\tt PyOSE} under the assumption of both bodies being black disks, and limb darkening being constant under the (small) area of the moon. The transit cord of both bodies over the stellar surface is calculated with the analytical model of \citet{2002ApJ...580L.171M}.

First, we simulated transits of a Jupiter-sized planet ($R_{\rm p}~=~R_{\rm Jup}$) with a Ganymede-sized test moon transiting a Sun-like star with $b~=~0$ over a range of possible planet-moon orbital separations ($1~\leq~r/R_{\rm p}~\leq~100$). Although the orbital separation ($a$) between the star and the planet-moon barycenter is irrelevant for our purposes, we chose $a=5.2$\,AU. The stellar radius is equal to the solar radius. Hence, $R_{\rm s} = 10\,R_{\rm p}$ and for $b=0$ the transit chord has a length of $2p = 20\,R_{\rm p}$, whereas for $b=0.5$ we have $2p = 17.4\,R_{\rm p}$.

Second, we simulated an Earth-Moon analog system ($R_{\rm p}~=~R_{\rm Ear}$), again for both $b~=~0$ and $b~=~0.5$, but now at $a~=~1$\,AU from a Sun-like host star. In the first case, the transit path had a length $2p = 218\,R_{\rm p}$ and in the latter case a length $2p = 189\,R_{\rm p}$. We also calculated the Hill sphere of the Earth as per $R_{\rm H} = a ( \, M_{\rm p} / (3 M_{\rm s}) \, )^{1/3}$, with $M_{\rm p}$ and $M_{\rm s}$ chosen as one Earth mass and one solar mass, respectively.

For each of the 100 test systems across the planet-moon distance interval that we investigated, we simulated 10,000 transits and computed the mean and standard deviation of $u$. The temporal resolution of each transit covered $1{,}000$ data points during planetary transit. In each transit, the initial moon phase was randomized. 



\section{Results}

\subsection{Jupiter-like exoplanets and Galilean-like exomoons}

In Fig.~\ref{fig:u_SNR_Jup}(a) we illustrate $u(r)$ for a hypothetical Sun-Jupiter-moon system. On the primary abscissa, $r$ is measured in units of the stellar radius and on the secondary abscissa, $r$ is measured in units of the planetary radius. The solid white line assumes a transit impact parameter $b=0$ following Eq.~\eqref{eq:u_solved}, and the dashed white line assumes $b=0.5$. The solid black line shows the outcome from our numerical transit simulations. For reference, the orbits of the Galilean moons are indicated with initials, with Io ($r/R_{\rm Jup}=6.1$), Europa ($r/R_{\rm Jup}=9.7$), Ganymede ($r/R_{\rm Jup}=15.5$), and Callisto ($r/R_{\rm Jup}=27.2$).

As a first observation, we find that the analytical framework (for $r>2p=20\,R_{\rm p}$) and the mean values after 10,000 numerical transit simulations per data point ($r/R_{\rm p}$) agree within about 1\,\%. The standard deviation obtained from the numerical simulations is quite substantial though and typically between 10\,\% and 20\,\%. Consequently, in real cases where just a few transits are observed, $u$ might differ substantially from the mean $u$ values that we present. That is, in fortunate (unfortunate) cases with just a few transits, most (very little) of the moon's in-transit data could be contaminated by the concomitant planetary transits.

Second, the corresponding values of $u$ for the Galilean moons are 14\,\% for Io, 20\,\% for Europa, 42\,\% for Ganymede, and 73\,\% for Callisto. These values, especially for the inner three Galilean moons, are astonishingly small and do not bode well for folding techniques of exomoon transit light curves.



In Fig.~\ref{fig:u_SNR_Jup}(b), we show $R=\sqrt{u}$, which corresponds to the reduced S/N due to the loss of the moon's in-transit data inside the planetary transits of the light curve. We find that moons in orbits similar to the Galilean moons preserve a S/N of about 38\,\% (Io), 45\,\% (Europa), 65\,\% (Ganymede), and 85\,\% (Callisto), respectively. For moons in wider orbits with $r > 35\,R_{\rm p}$, we find that $R$ increases beyond 90\,\%. As a side note, and not shown in any of our figures, we find that for Titan around Saturn, which is at $r=21.3\,R_{\rm p}$, Eq.~\eqref{eq:u_solved} yields $u=63\,\%$ and $R=79\,\%$. These values are more akin to those of Callisto than to those of Ganymede.

The measurement by K21 in a hypothetical planet-exomoon system in an extremely wide orbit ($r/R_{\rm Jup}=44.3$), based on 24 simulated transits (for $b=0.5$) with Kepler-like space-based transit photometry, is labeled as well in Fig.~\ref{fig:u_SNR_Jup}(b). While this previous measurement suggested $R \approx 82\,\%$, we find $R \approx 93\,\%$. That said, our 10,000 numerical transit simulations at $r/R_{\rm Jup}=44$ exhibit a standard deviation of 27\,\%, so that the previous measurement is well within the statistical fluctuations.

\subsection{Earth-like exoplanets and Moon-like moons}

In Fig.~\ref{fig:u_SNR_Ear} we shift our focus to the Earth-Moon system. The Moon's orbital separation from Earth is labeled with an "M" at $r/R_{\rm Ear}=60.3$, at which distance $u = 18\,\%$ and $R = 42\,\%$. Any moon orbit beyond about $235\,R_{\rm Ear}$, or $2.15\,R_{\rm Sun}$, in Fig.~\ref{fig:u_SNR_Ear} is beyond the Earth's Hill sphere and therefore dynamically unstable. We note that the Moon is at a distance of about 26\,\% of the Earth's Hill sphere and that prograde moons can only be stable out to about $50\%$ of $R_{\rm H}$ \citep{2006MNRAS.373.1227D}, where we find $u = 64\,\%$ and $R = 80\,\%$. This means that any exomoon's transit S/N around an Earth-mass planet can be expected to be reduced to 80\,\%, and typically much less. Again, this finding and the low values of $u = 18\,\%$ and $R = 42\,\%$ for the Earth's Moon are astonishingly small and reveal a significant caveat of folding techniques of exomoon transit light curves.

In comparing Figs.~\ref{fig:u_SNR_Jup} and \ref{fig:u_SNR_Ear}, we confirm that the white and black solid curves therein are virtually identical when scaled in units of the stellar radius, except for slight statistical variations in the numerical simulations. We also notice that the fluctuations of the standard deviation are much more pronounced in Fig.~\ref{fig:u_SNR_Jup} than in Fig.~\ref{fig:u_SNR_Ear}. This is due to mutual planet-moon eclipses, which are much more often and longer lasting for moons around the Jupiter-sized test planet than for moons around the Earth-sized test planet. First, at a given distance from the planet, the moon's orbital velocity is larger around the more massive test planet, increasing the frequency of eclipses. And second, the larger planet simply yields a higher eclipse probability.

As a side note, the Hill radius of a Jupiter-mass planet at 1\,AU from a Sun-like star is $148\,R_{\rm Jup}$, or $14.7\,R_{\rm Sun}$. Moreover, we note that in the Earth-Moon test system, the condition $r > 2p$ is only met for $b > 0.96$, so that an uncontaminated fraction of the Moon's OSE probability density (blue area in Fig.~\ref{fig:OSE}) can occur across the solar-type host star.


\subsection{The $2/\pi$ limit of uncontaminated exomoon transit data}

Finally, as a special case for moons in orbits that are closer than the transit chord, Eq.~\eqref{eq:u_solved} reveals a previously unknown, fundamental limit to the maximum fraction of the moon's in-transit data that is uncontaminated by planetary transits. In the limit of planet-moon orbital separations that are as wide as the transit chord across the star ($r=2p$) and with $(R_{\rm p}/r)^2$ being typically on the percent or permille level for the large moons of the Solar Systems, Eq.~\eqref{eq:u_solved} collapses to 

\begin{equation}
\lim_{r \rightarrow 2p}(u) = 2/\pi \sqrt{1-(R_{\rm p}/r)^2} \approx 2/\pi \approx 63.7\,\% \ .
\end{equation}

\noindent
For moons in circumplanetary orbits that are closer than the transit chord ($r<2p$), we derived $u<2/\pi \approx 63.7\,\%$. In other words, a maximum fraction of $u = 2/\pi$ of the moon's in-transit data can be observed out of the planetary transits in these systems, given a sufficiently large number of transits.

\section{Discussion}

As for the difference between 93\,\% of the maximal S/N as per our analytical solution and 82\,\% measured by K21, we note the following. (1) The algorithm of K21 cuts out some moon transit data outside the planetary transits to prevent the planetary transits from leaking into the phase-folded data. As a consequence, more data are lost than we considered in our model, in which the loss of data is restricted to the moon's in-transit data during planetary transits only. (2) K21 compares a box-fitting algorithm with a full photodynamical model, the latter of which likely used a realistic transit model. This induces a systematic reduction of the relative S/N estimate compared to our purely theoretical considerations.

In studying Fig.~\ref{fig:u_SNR_Jup}, we find that the value of $R \approx 82\,\%$ by K21 is plausible for this particular example, but likely not representative of extrasolar moons in general. It is plausible because it is well within the standard deviation of our numerical simulations of a star-planet-moon system similar to that of K21. It is likely not representative though because, first, large moons in such extremely wide orbits of about $44\,R_{\rm Jup}$ are not known in the Solar System. And second, a moon in such a wide circumplanetary orbit at the planetary Hill radius would hardly be stable due to the gravitational perturbations caused by the star. In fact, the maximum distance of a prograde moon is about half the planetary Hill radius.

\section{Conclusions}

Exomoons in orbits reminiscent of the Galilean moons around Jupiter, or similar to Saturn's moon Titan, or akin to the Earth's moon have large parts of their in-transit data contaminated by the concomitant planetary transits. Specifically, we find that the S/N of the exomoon signal is reduced due to contamination from the planetary transits to 38\,\% for Io, 45\,\% for Europa, 65\,\% for Ganymede, 85\,\% for Callisto, 79\,\% for Titan, and 42\,\% for the Earth's Moon when compared to the S/N of a fully consistent photodynamical model that includes the entire exomoon transit data and in particular those during planetary transit. As a consequence for exomoon searches, dismissing any piece of data from the light curve that is contaminated by the planetary transit leads to a substantial loss of S/N for realistic moons.

The gain in speed for any exomoon transit search algorithm that ignores the planetary in-transit data comes at the heavy price of losing a substantial fraction of what is supposedly a tiny signal in the first place. Photodynamical modeling that includes both the planet's and the moon's transit data as well as possible planet-moon eclipses is probably required for the detections of exomoons.

\begin{acknowledgements}
The authors thank an anonymous referee for a swift and collegial report. RH acknowledges support from the German Aerospace Center (Deutsches Zentrum f\"ur Luft- und Raumfahrt) under PLATO Data Center grant 50OO1501.
\end{acknowledgements}

\bibliographystyle{aa}
\bibliography{literature}

\appendix



\end{document}